\documentclass[aps,groupedaddress,showpacs]{revtex4}
\usepackage{amsmath,amscd,amssymb,epsfig,graphics,amsbsy}
\begin{document}
\bibliographystyle{apsrev}
\title{Fractional diffusion modeling of ion channel gating}
\author{Igor Goychuk}
\email[]{goychuk@physik.uni-augsburg.de}
\author{Peter H\"anggi}
\affiliation{Institute of Physics, University of Augsburg,
Universit\"atsstr. 1, D-86135 Augsburg, Germany}
\date{\today}


\begin{abstract}
An anomalous diffusion model for  ion channel gating is put forward.
This  scheme is able to describe non-exponential, power-law like
distributions of residence time intervals in several types of ion
channels. Our method presents a generalization of the discrete
diffusion model by Millhauser, Salpeter and Oswald [Proc. Natl.
Acad. Sci. USA 85, 1503 (1988)] to the case of a continuous,
anomalous slow conformational diffusion. The corresponding
generalization is derived from a continuous time random walk
composed of  nearest neighbor jumps which in the scaling limit
results in a fractional diffusion equation. The studied model
contains three parameters only: the mean residence time, a
characteristic time of conformational diffusion, and the index of
subdiffusion. A tractable analytical expression for the
characteristic function of the residence time distribution  is
obtained. In the limiting case of normal diffusion, our prior
findings [Proc. Natl. Acad. Sci. USA 99,
3552 (2002)] are reproduced. Depending on the chosen parameters, the
fractional diffusion model exhibits a very rich behavior of the
residence time distribution with different characteristic
time-regimes. Moreover, the corresponding autocorrelation function
of conductance fluctuations displays nontrivial features.  Our
theoretical model is in good agreement with experimental data for
large conductance potassium ion channels.
\end{abstract}

\maketitle

\section{Introduction}

Ion channels are complex membrane proteins which provide
ion-conducting, nanoscale pores in the biological membranes
\cite{Hille}. These proteins undergo spontaneous conformational
dynamics resulting in stochastic intermittent events of opening and
closing the pore -- the so-called gating dynamics. It can be
 described by following kinetic scheme:
\begin{eqnarray}\label{scheme}
C \begin{array}{c}  k_o \\[-0.1cm] \longrightarrow
\\[-0.2cm] \longleftarrow \\[0cm] k_c \end{array} O  \;.
\end{eqnarray}
As it stands, this scheme describes Markovian stochastic transitions
between the closed state (C) and the state open (O) of an ion
channel which can  fully be characterized by the opening rate,
$k_o$, and the closing rate, $k_c$. From a trajectory description of
the observed two-state gating process, these transitions can be
characterized by the residence time distributions (RTDs) of open and
closed time intervals, $\psi_o(\tau)=k_c\exp(-k_c\tau)$ and
$\psi_c(\tau)=k_o\exp(-k_o\tau)$, respectively.

The invention of
patch clamp technique \cite{Neher} marked the beginning of a new
era: detailed experimental studies of the statistics of such
stochastic trajectory realizations have been rendered possible.
These experimental investigations, however, also reveal the fact
that the distributions of the residence time intervals are typically
not exponential. This in turn implies that the corresponding
 {\it observed} two-state dynamics of current fluctuations
 is {\it not} Markovian. Any such non-exponential distribution can however
approximately be fitted
by a sum of (sometimes many) exponentials, e.g.
\begin{equation}\label{fit}
\psi_c(t)=\sum_{i=1}^{N}c_i\lambda_i\exp(-\lambda_i t),
\end{equation}
with weights $c_i$ obeying $\sum_{i=1}^{N}c_i=1$. The rationale
behind this fitting procedure is the assumption that the
corresponding state consists of $N$ discrete substates, separated by
potential barriers. This method constitutes the working tool for the
majority of  molecular physiologists in interpreting their
experimental data within a discrete Markovian scheme consisting of
many (sub)states \cite{Colquinon}. The addition of new states (or
new configurational dimensions in the continuous case) is a well
known formal method to unravel a low-dimensional non-Markovian
stochastic dynamics via its embedding into  a
Markovian dynamics of higher dimensionality. The problem with such a
methodology is, however, that the number of substates needed to fit
the experimental data can depend on the experimental conditions. For
example,  the experimental gating dynamics of a Shaker potassium
channel has been successfully described by a sequential 8-state
Markovian scheme with 7 closed states for a fixed value of
temperature about $T=20\;^oC$ \cite{Bezanilla94}. However, to
describe the experimental data over a small extended temperature
regime between $10-20\;^oC$ already necessitates to add three
additional closed substates \cite{Rodriguez98}.

For several types of
ion channels the RTDs can alternatively be fitted in terms of non-exponential
distributions such as a stretched exponential $\psi(\tau)\propto
-\frac{d}{d\tau}\exp[-(\nu\tau)^{\alpha}]$ \cite{Liebovitch},
or by a power law
$\psi(\tau)\propto \tau^{-\beta}$ \cite{Millhauser,Sansom} with a
few parameters only. The case of a power law coefficient near
$\beta=3/2$ can be described with normal conformational diffusion
over many degenerate substates
\cite{Millhauser,Condat,NadlerStein,PNAS02,Physica03}.
Such diffusion models \cite{Millhauser,Condat,
NadlerStein,PNAS02,Physica03,Lauger,Levitt,Shirokov,Sigg}
present an alternative to
the standard discrete state Markovian modeling which can be
considered also as a complementary one \cite{Sigg}.
It should be noted, however,
that exponents $\beta$ different from a normal diffusion behavior
have been detected experimentally as well
\cite{gramicidin,Gorz,Mercik}. Therefore, it is of
prominent importance to generalize the normal diffusion model to
the case of anomalous diffusion: this objective is at the heart of
the following discussion.

\section{Modeling anomalous ion channel gating}

Ion channels are protein complexes with an intrinsically
hyperdimensional conformational space. Such macromolecular complexes
can possess several macroconformations corresponding to different
functional states of the ion channel. These macroconformations are,
however, not static, but rather present dynamical structures
featuring multiple conformational substates and undergoing a
corresponding intraconformational dynamics which can be of the
diffusion type, as it is well studied for one of the simplest
proteins -- myoglobin \cite{Frauenfelder,FM2000,PN2002}. It
particular, for myoglobin it has been shown experimentally  that the
diffusive motions become increasingly important above some critical
temperature and can serve as lubricants  for macroconformational
changes, see e.g. in Ref. \cite{PN2002}. There is no convincing
reason to believe that ion channels, which are structurally much
more complex proteins, are very different from myoglobin in this
respect. Therefore, the conformational diffusion should be present
and can be important for the gating dynamics of ion channels as
well. Diffusion models of ion channel gating put emphasis on this
fundamental feature of protein dynamics. In the case of ion channels
with one open and one closed macroconformations the whole
conformational space can be divided into the two hyperdimensional
subspaces corresponding to the {\it open} macroconformation and to
the {\it closed} macroconformation, respectively. These subspaces
present generally some complex manifolds consisting of domains of
attraction in the conformational space which are separated by
potential barriers. Such domains of spatial localization can be
associated with discrete states in the discrete state diffusion
modeling of the gating dynamics. They can possess but a complex
intrinsic structure reflecting the intrinsic multidimensionality of
the problem. For this  reason, such states can be trapping states
with a non-exponential distribution of the residence times spent in
the corresponding domains of attraction. On the contrary, the
Markovian assumption within a discrete state modeling involves the
assumption on strictly exponential distribution of the residence
times in such discrete states. This presents at best an
approximation only. It is at the heart of modeling the gating
dynamics with appropriate Markovian kinetic schemes. Moreover, some
parts of conformational potential landscape can lack the presence of
pronounced barriers and some domains of attraction can be flat
providing valleys in the corresponding potential landscape. 
This provides an ideal starting point for a
diffusion modeling of the gating dynamics.

The simplest diffusion model \cite{Millhauser,Condat} assumes the
existence of such a one-dimensional valley wherein the diffusion
 is modeled by {\it sequential} transitions among a large number of
discrete substates. Alternatively, a continuous diffusion modeling
can be applied \cite{Levitt,NadlerStein,Shirokov,PNAS02}. Note also
that multiple pathways into the open state can be accounted for within
a one-dimensional reaction coordinate modeling by allowing for
transitions with non-nearest neighbor jumps. Such complexity will not be
addressed within our sequential diffusion model. At one edge of
this valley the transition into the open state can occur via the
single route \cite{Millhauser,Condat,NadlerStein}. Such a modeling
is appropriate when the correlations between the subsequent closed
and/or open residence time intervals are absent. The 
multiple transitions could also induce some correlation between
the durations of time intervals. 
Such effects will be  neglected in a zero order
approximation (renewal assumption for the observed two-state
conductance fluctuations). The existence of time-correlations
between the durations of time intervals (non-renewal gating
dynamics) would indicate the presence of more than one route between
the manifolds of open and closed states (see, e.g., in Ref.
\cite{Neher}, pp. 440-441). Such correlations can also be accounted
for in the diffusion models \cite{Levitt,Shirokov}. The
incorporation of such correlations is, however, beyond the scope of
our present work which presents an  anomalous diffusion
generalization of the Markovian modeling used in  Ref.
\cite{Millhauser,Condat,NadlerStein,PNAS02}.


Let us start from a continuous time random walk (CTRW)
\cite{MontrollWeiss,LaxSher,Hughes,Shlesinger,Weiss} generalization
of the discrete state diffusion model by Millhauser {\it et al.}
\cite{Millhauser}, see Fig. 1. It is assumed that the manifold of
closed substates consists of N states; namely, the states from $j=2$
to $j=N-1$ (``diffusion states'') are identical and characterized by
identical residence time distributions $\psi_j(\tau)=\psi(\tau)$,
i.e. the channel stays in the corresponding state $j$ for a random
time interval $\tau$ distributed in accordance with $\psi(\tau)$,
and performs at the end of every time interval a jump $ j
\rightarrow j \pm 1$
 with  probability $p_{j+1,j}=p_{j-1,j}=1/2$ either to
the left, or to the right neighboring state, respectively.
If the RTD is exponential, i.e.
$\psi(\tau)=2\kappa\exp(-2\kappa\tau)$,
then the standard Markovian rate description
with rate $\kappa$
is recovered. The boundary
state $j=N$ possesses a different RTD $\psi_N(\tau)$
which in the Markovian case reads $\psi_N(\tau)=\kappa\exp(-\kappa\tau)$
(transitions occur always with
the probability one, $p_{N-1,N}=1$, to the  state $j=N-1$).
Furthermore, from the state $j=1$ the channel can undergo a
transition into its open state $j=0$ with the rate $\gamma$, or make
transition into the manifold of conformational diffusion substates
with the rate $\kappa$. For this state, the corresponding RTD reads
$\psi_1(\tau)=(\gamma+\kappa)\exp[-(\gamma+ \kappa)\tau]$ and the
transition probabilities are $p_{01}=\gamma/(\gamma+\kappa)$ and
$p_{21}=\kappa/(\gamma+\kappa)$. The dynamics of state occupancies
$p_j(t)$ is described by the generalized master equation (GME) due
to Kenkre, Montroll and Shlesinger \cite{Kenkre73,Weiss83} and its
generalization \cite{BurstZharTem,remark}.
\begin{figure}
\begin{center}
\epsfig{file=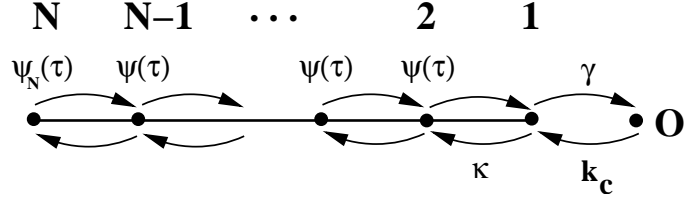,width=0.5\textwidth}
\end{center}
\caption{Sketch of the CTRW generalization of the (discrete) diffusion
model of ion channel gating. }
\label{Fig1}
\end{figure}
The corresponding dynamics reads (with an  initial preparation in
some state $j_0$, $p_{j_0}(0)=1$):
\begin{eqnarray}
\dot p_j(t)& = &\int_0^{t}K(t-t')[p_{j-1}(t')+p_{j+1}(t')
-2p_j(t')] dt', \;j=\overline{3,N-2},\label{master}\\
\dot p_N(t)& = & \int_0^{t}K(t-t')p_{N-1}(t')dt'-
\int_0^{t}K_N(t-t')p_{N}(t')dt', \label{second}\\
\dot p_{N-1}(t)& = &\int_0^{t}K(t-t')[p_{N-2}(t') \label{5th}
-2p_{N-1}(t')] dt'+\int_0^{t}K_N(t-t')p_{N}(t')dt',\\
\dot p_2(t) & = &\int_0^{t}K(t-t')[p_{3}(t')
-2p_2(t')] dt' + \kappa p_1(t),\label{6th} \\
\dot p_1(t) & = & \int_0^{t}K(t-t')p_{2}(t')dt'-(\kappa+\gamma)
p_1(t) + k_c p_0(t),\label{7th} \\
\dot p_0(t) & = &- k_c p_0(t) +\gamma p_1(t),
\end{eqnarray}
with the kernels $K(t)$ and $K_N(t)$ defined through their
Laplace-transforms
\begin{eqnarray}\label{kernel}
\tilde K(s) &= &\frac{1}{2}\frac{s\tilde\psi(s)}
{1-\tilde\psi(s)},\nonumber \\
\tilde K_N(s)& = & \frac{s\tilde\psi_N(s)}{1-\tilde\psi_N(s)},
\end{eqnarray}
where $\tilde\psi(s)$ and $\tilde \psi_N(s)$ are the
Laplace-transforms of $\psi(\tau)$ and $\psi_N(\tau)$,
respectively.
These kernels can be derived due to the following
simple consideration. Let us prepare the diffusing particle at $t=0$ in the
state $j=2...N-1$ and let it make transitions
to the neighboring states without return. Then,
the survival probability $\Phi(t)=\int_{t}^{\infty}\psi(\tau)d\tau$
in the state $j$ (which should be identified under such conditions with
$p_j(t)$) is governed
obviously by the following equation
$\dot \Phi(t)=-2\int_0^t K(t-t')\Phi(t')dt'$ with $\Phi(0)=1$.
It can be easily solved by the Laplace-transform method to yield
the first equation in Eq. (\ref{kernel}) by taking into account
$\tilde \psi(s)=1-s\tilde\Phi(s)$. The second kernel $K_N(t)$ is obtained
along the same lines.

The RTD of the open state is readily obtained,
i.e. $\psi_o(\tau)=
k_c\exp(-k_c\tau)$. In order to calculate the RTD of the set of closed
states  one starts out from $p_1(0)=1$ (the channel has
been just closed) to obtain  the survival probability
$\Phi_c(t)=\sum_{j=1}^{N}p_j(t)$ with
the boundary condition that the state $j=0$ is absorbing.
This latter condition  is realized by setting  formally $k_c\to 0$.
The corresponding RTD then follows from $\psi_c(\tau)=-d\Phi_c(\tau)/d\tau$.
The total population of the closed state
$p_c(t)=\sum_{j=1}^{N}p_j(t)$ obeys (not allowing for
the backward transition, $k_c\to 0$):
\begin{eqnarray}\label{boundary1}
\dot p_c(t)=-\gamma p_1(t).
\end{eqnarray}
This must be used as the proper boundary condition (it yields a radiation
boundary via continuity equation in the scaling limit,
see below) to calculate
$\psi_c(\tau)$ \cite{KampenBook,Goel1974}.

Next let us make the Ansatz that $\psi(\tau)=-\frac{d}{d\tau}
E_{\alpha}(-(2\kappa\tau)^{\alpha})$, where
$E_{\alpha}(z):=\sum_{0}^{\infty}z^n/\Gamma(\alpha n+1)$ is the
Mittag-Leffler function. It is defined via a generalization of the
Taylor series expansion of the exponential function,
$E_1(z)=\exp(z)$ and $\Gamma(z)$ is the standard Gamma-function.
In other words, the corresponding
survival probability  in the state just occupied \cite{Cox},
$\Phi(\tau)=\int_{\tau}^{\infty}
\psi(t)dt$ is given by $\Phi(\tau)=E_{\alpha}(-(2\kappa\tau)^{\alpha})$.
This corresponds to the Cole-Cole relaxation law, i.e. the
relaxation law which yields the Cole-Cole susceptibility
\cite{Cole,Hilfer1}.
The Laplace-transform of $\Phi(\tau)$
reads $\tilde \Phi(s)=s^{\alpha-1}/
[s^\alpha+(2\kappa)^\alpha]$
\cite{GorenfloMainardi,MetzlerKlafter} and by use of the
relation, $\tilde \psi(s)=1-s\tilde \Phi(s)$,
one obtains $\tilde \psi(s)=
(2\kappa)^\alpha/[s^\alpha+(2\kappa)^\alpha]$.
This particular choice of RTD interpolates between the initial stretched
exponential (Weibull) distribution \cite{Cox} (which corresponds to the
Kohlrausch relaxation law;
``stretched exponential'' refers here to the survival probability
$\Phi(\tau)$, with $\psi(\tau)\propto 1/\tau^{1-\alpha}$ at $\tau \to 0$)
and the asymptotic
long time power law $\psi(\tau)\propto 1/\tau^{1+\alpha}$. It
yields anomalously slow diffusion, $\langle \delta x^2(t)\rangle
\propto t^{\alpha}$ \cite{MetzlerKlafter}. For example,
such an anomalous diffusion  is measured
experimentally, along with the RTD in trapping domains exhibiting
the corresponding power law,  for colloidal particles in cytoskeleton
actin networks of biological cells \cite{Wong}
and in living cells \cite{Tolic}. These latter
experimental results offer a clear clue for understanding the
results on virus diffusion in infected cells \cite{Brauchle} -- that
is an observed anomalously slow diffusion in actin networks combined
with active directional trafficking of viruses by molecular motor
proteins \cite{Brauchle}.  For the discussed form of RTD, $\tilde
K(s)=(2\kappa)^\alpha s^{1-\alpha}/2$ and the fractional master
equation follows  exactly from Eq. (\ref{master}) as a particular
case of the GME. It reads explicitly,
\begin{eqnarray}\label{fracmaster}
\dot p_j(t) = \frac{1}{2}(2\kappa)^\alpha\sideset{_0}{_t}
{\mathop{\hat D}^{1-\alpha}}[p_{j-1}(t')+p_{j+1}(t')
-2p_j(t')],
\end{eqnarray}
where $\sideset{_0}{_t}{\mathop{\hat D}^{1-\alpha}}
(...)=\frac{1}{\Gamma(\alpha)}\frac{\partial}{\partial
t}\int_{0}^{t}\frac{(...)dt'}{(t-t')^{1-\alpha}}$ is the
integro-differential  operator of fractional derivative introduced
by Riemann and Liouville, see Refs.
\cite{GorenfloMainardi,MetzlerKlafter,Sokolov,Hilfer} for reviews
and further references. In the case of a two-state dynamics a
similar fractional master equation was obtained in
\cite{SokolovMetzler,Stan}.
 Note that the fractional master
 equation (\ref{fracmaster})
 presents in fact a conventional
generalized master equation  being nonlocal in time. The
introduction of a fractional time derivative in the generalized
master equation of CTRW is nothing but a shorthand notation which
corresponds to a specific choice of the RTD. The importance of this
equation lies in the fact that it can serve as a useful mathematical
tool to model  anomalously slow diffusion. The physical origin of this
diffusion can be attributed to very broad residence time
distributions on the sites of particle localization with diverging
mean residence time \cite{MontrollWeiss,Hughes,Shlesinger}. In
practice this  implies that the corresponding mean residence
time is exceedingly large as compared with the characteristic time
scale of anomalous diffusion in the given domain of a finite size.
As a consequence, the  approximation
with an infinite mean residence time becomes physically justified.

Likewise,  with $\psi_N(\tau)= -\frac{d}{d\tau}
E_{\alpha}(-(\kappa\tau)^{\alpha})$, Eq. (\ref{second}) takes on the
form
\begin{eqnarray}
\dot p_N(t) =  \frac{1}{2}(2\kappa)^\alpha\sideset{_0}{_t}
{\mathop{\hat D}^{1-\alpha}}p_{N-1}(t')-\kappa^
\alpha\sideset{_0}{_t}{\mathop{\hat D}^{1-\alpha}}
p_{N}(t'). \label{secondfrac}
\end{eqnarray}
The remaining  equations (\ref{5th}), (\ref{6th}), and (\ref{7th})
involving the memory
kernel can readily be rewritten in a similar form upon use of the
notation of the fractional time derivative.

It is worth noting that Eq. (\ref{fracmaster}) and Eq. (\ref{secondfrac})
can equivalently be brought into a form with the fractional Caputo
derivative
$D_{*}^{\alpha}(...):=\frac{1}{\Gamma(1-\alpha)}
\int_0^t\frac{\frac{\partial}{\partial t'}(...)d t'}{(t-t')^\alpha}$
\cite{GorenfloMainardi}
in the left hand side. Namely, Eq. (\ref{fracmaster}) becomes
\begin{eqnarray}\label{fracmaster3}
D_{*}^{\alpha} p_j(t') = \frac{1}{2}(2\kappa)^\alpha[p_{j-1}(t)+p_{j+1}(t)
-2p_j(t)]\;.
\end{eqnarray}
Such a fractional master equation has been first derived from CTRW in Ref.
\cite{HilferAnton}. This form is, however, clearly not appropriate for
a further use in the
studied case, where the normal rate transitions are also present. The reason
is that
the Eqs. (\ref{6th}), (\ref{7th}) would get a form mixing
the fractional Caputo derivative
and the fractional
Riemann-Liouville
integral in left and right hand sides of the
corresponding equations. This fact establishes a clear
advantage for the use of the Riemann-Liouville fractional derivative in
a case where normal rate processes
are also present.

\subsection {Scaling limit to a fractional diffusion equation}

Let us perform next a continuous limit: namely, assuming the
distance $\Delta x$ between neighboring sites we introduce the
conformational coordinate $x:=-j\Delta x$ which models the manifold
of closed diffusion substates.  The following continuous limit is
assumed: Let $\Delta x\to 0$, $N\to \infty$, $\kappa\to \infty$
whereas keeping $K_{\alpha}:=\frac{1}{2}(2\kappa)^\alpha(\Delta
x)^2$ and the diffusion ``length'' $L:=N\Delta x$ constant. By use
of the expansion: $p_{j\pm 1}(t)/\Delta x:=P(-[j\pm 1]\Delta x,t)\approx
P(x,t)\mp \frac{\partial P(x,t)}{\partial x}\Delta x+
\frac{1}{2}\frac{\partial^2 P(x,t)}{\partial x^2}(\Delta x)^2$, in
(\ref{master}) we arrive at the following fractional diffusion
equation in continuous state space, i.e.,
\begin{eqnarray}\label{fracdif}
\frac{\partial P(x,t)}{\partial t}=K_{\alpha}  \sideset{_0}{_t}
{\mathop{\hat D}^{1-\alpha}}
\frac{\partial^2  P(x,t')} {\partial x^2},
\end{eqnarray}
where $K_{\alpha}$ is the diffusion constant of anomalously slow
diffusion.
The fractional diffusion equation (\ref{fracdif}) presents a
non-Markovian diffusion equation \cite{Weiss} with a particular
choice of an integral kernel that accounts for the residence time
distributions following the Cole-Cole relaxation law
\cite{Cole,Hilfer1} in a corresponding formulation in terms of a
discrete CTRW.
The fractional diffusion equation (\ref{fracdif})
assumes the form of a   continuity equation,
\begin{eqnarray}\label{contin}
\frac{\partial P(x,t)}{\partial t}=-\frac{\partial J(x,t)}{\partial x}
\end{eqnarray}
where the probability flux $J(x,t)$ becomes modified due to the
fractional time-derivative,
\begin{eqnarray}\label{flux}
J(x,t)=-K_{\alpha}  \sideset{_0}{_t}
{\mathop{\hat D}^{1-\alpha}}
\frac{\partial  P(x,t')} {\partial x} \;.
\end{eqnarray}
Our derivation of the fractional diffusion equation from the CTRW
complements previous studies \cite{MetzlerKlafter,Metzler}; it is
rather simple and does not require  jumps with a variable step
length beyond nearest neighbors. For this very reason, no
overflights of the boundaries occur that are  possible otherwise.
This observation is of crucial importance in determining the
physically correct boundary conditions
\cite{PRA84,Balakrishnan88}. This also means that the boundaries are
strictly local in space. The given derivation removes any open query
about the space-locality of boundary conditions for the fractional
diffusion equation. After the integration of the continuity equation
from $x=-L$ to $x=0$, one deduces that the decrease of the total
probability of the closed state manifold (the survival probability),
$p_c(t)= \int_{-L}^0 P(x,t)dx$ occurs due to the probability flux on
the boundary. Accordingly, we will replace the original discrete
master equation in space  by its corresponding  fractional
diffusion equation with the following boundary conditions as they
emerge from the original problem. The boundary $x=-L$ is a
reflecting one, obeying :
\begin{eqnarray}\label{reflect}
J(-L,t)=0\;.
\end{eqnarray}
The boundary at $x=0$ is radiative. Setting $p_1(t)\approx \Delta x
P(0,t)=L P(0,t)/N$ and using Eq. (\ref{boundary1}) one finds
\begin{eqnarray}
J(0,t)= \gamma L P(0,t)/N.
\end{eqnarray}
We additionally use here the specific scaling limit: $\gamma\to
\infty$, $N\to \infty$ with $\langle \tau_c\rangle=N/\gamma$ being
held fixed. This quantity possesses the meaning of the mean
residence time in the closed state manifold, see below. The
radiation boundary acquires then the explicit form
\begin{eqnarray}\label{radbound}
J(0,t)= L P(0,t)/\langle \tau_c\rangle.
\end{eqnarray}
Our fractional diffusion modeling for the RTD of closed time
intervals thus has three parameters only: (i) the mean residence
time $\langle \tau_c\rangle$, (ii) the characteristic time of
conformational diffusion, i.e.  $\tau_D:=(L^2/K_\alpha)^{1/\alpha}$
and (iii) the power law index of anomalous diffusion $\alpha$. In
the case of normal diffusion, i.e. $\alpha=1$, the special
scaling limit used here can also  be justified from a Kramers
approach \cite{RMP90} to the gating
problem  \cite{PNAS02,Physica03}. Whereas the
solution of the Kramers approach   in Ref. \cite{PNAS02}
in addition yields also an analytical expression for $\langle
\tau_c\rangle$, which reproduces the experimental crossover
 behavior from an
exponential-to-linear voltage dependence due to Hodgkin and
Huxley \cite{HH},
the model here treats $\langle
\tau_c\rangle$ as one of the phenomenological
parameters. It is also worthwhile to  remark that the boundary
condition does {\it not} contain the index of anomalous diffusion
$\alpha$. Formally it remains the same as for normal diffusion. The
flux expression (\ref{flux}) is, however,  not local in time.
Moreover,  the r.h.s. of (\ref{radbound}) does  not contain the
fractional derivative in time. This feature is in accord with the
original discrete model where the last, final transition into the
open state is given by an ordinary  rate transition.
The analogy with the Kramers model of
Ref. \cite{PNAS02} is that the diffusion in
the domain of voltage-dependent states
(cf. \cite{PNAS02,Physica03}), which becomes a thin boundary layer
in the considered scaling limit,
remains {\it normal}. This
justifies well the use of boundary condition (\ref{radbound}) in our
fractional diffusion
model which now is completely formulated.

\subsection{Characteristics of the residence time distribution}
To obtain the distribution of closed residence times $\psi_c(\tau)$
one needs to solve first the fractional diffusion equation
(\ref{fracdif}) with the boundary conditions (\ref{reflect}) and
(\ref{radbound}) and the initial condition $P(x,0)=\delta(x-x_0)$
with $x_0\to 0_{-}$. The survival probability $\Phi_c(t)$ follows as
the integral of the solution over the spatial variable and
subsequently the corresponding RTD follows  as the negative time
derivative of the survival probability. This task has been achieved
by use of   the Laplace-transform method. The details of the
derivation are outlined in the Appendix. The final result for the
Laplace-transformed RTD of  closed time intervals then reads:
\begin{eqnarray}\label{mainres}
\tilde \psi_c(s) = \frac{1}{1 + s\langle \tau_c\rangle
g_{\alpha}(s\tau_D)}  \;\;,
\end{eqnarray}
where an auxiliary function
 \begin{eqnarray}
g_{\alpha}(z) = \frac{ \tanh[z^{\alpha/2}]}{z^{\alpha/2}}
\end{eqnarray}
has been introduced. For $\alpha=1$, this result reduces to one for
normal diffusion in Ref. \cite{PNAS02,Physica03}. Moreover, since
$g_{\alpha}(z)=1+o(z)$ for small $z$, one can readily see from Eq.
(\ref{mainres}) that $\langle \tau_c\rangle$  indeed has the meaning
of a mean residence time, $\langle \tau_c\rangle:=
\int_0^{\infty}\tau\psi_c(\tau)d\tau=- \frac{d\tilde
\psi(s)}{ds}|_{s=0}$. Note also that  the second moment of RTD
diverges, $\langle
\tau_c^2\rangle:=\int_0^{\infty}\tau^2\psi_c(\tau)d\tau \to\infty$
for all $\alpha<1$ (anomalous diffusion). Furthermore, if
$\tau_D=0$, then the closed time distribution becomes strictly
exponential, i.e. $\psi_c(\tau)= \exp(-\tau/\langle \tau_c
\rangle)/\langle \tau_c \rangle$ and the simplest two-state
Markovian model of the ion channel gating is reproduced with the
opening rate $k_o=\langle \tau_c \rangle^{-1}$. In general, the
expression (\ref{mainres}) cannot be inverted to the time domain
exactly; its different characteristic regimes, however, can be
discussed analytically.

In proceeding, let us consider first the limit of a large
conformational diffusion time $\tau_D\gg\langle \tau_c\rangle$. Then
(by use of  the large-$z$ asymptotic behavior of $g(z)\sim
z^{-\alpha/2}$), we have
\begin{eqnarray}\label{asym1}
\tilde \psi_c(s) \approx
\frac{1}{1+ (s\tau_0)^{1-\alpha/2}}
\end{eqnarray}
with $\tau_0:=\tau_D\Big (\frac{\langle \tau_c\rangle}{\tau_D}\Big)^
{1/(1-\alpha/2)}$.
The inversion of Eq. (\ref{asym1}) is given as
$\psi_c(\tau)=-d\Phi_c(\tau)/d\tau$ in terms of
the survival probability
\begin{eqnarray}\label{asym1direct}
\Phi_c(\tau) =
 E_{1-\alpha/2}\Big[-
 \Big(\frac{\tau}{\tau_0}\Big)^{1-\alpha/2}\Big] \;,
\end{eqnarray}
which is expressed through the Mittag-Leffler function $E_{\alpha}(z)$
and corresponds to the Cole-Cole relaxation law \cite{Cole,Hilfer1}.
Because $E_{1/2}(-z^{1/2})=e^z{\rm erfc}(z^{1/2})$ \cite{MetzlerKlafter},
where ${\rm
erfc}(z)$ is the complementary error function,
the solution of the normal diffusion problem
in Ref. \cite{NadlerStein} for the initial and intermediate time
evolution  regimes is
reproduced from Eq. (\ref{asym1direct}). For $\tau \ll\tau_0$, Eq.
(\ref{asym1direct}) behaves
as a stretched exponential \cite{MetzlerKlafter},
\begin{eqnarray}\label{asym2}
\Phi_c(\tau) \approx
 \exp\Big[-\frac{1}{\Gamma(2-\alpha/2)}
 \Big(\frac{\tau}{\tau_0}\Big)^{1-\alpha/2}\Big] \;.
\end{eqnarray}
This dependence (\ref{asym2}) corresponds in the language
of time-dependent rates $k_o(\tau):=-\frac{d}{d\tau}\ln[\Phi_c(\tau)]$
(used in the renewal theory)
\cite{Cox,Kampen79,Liebovitch} to
$k_o(\tau) \propto \tau^{-\alpha/2}$.
Such a time-dependent rate of
recovery from inactivation  has been measured with $\alpha=1$ for a
sodium ion channel in Ref. \cite{Toib}.
In the limit $\tau\to 0$, Eq. (\ref{asym2}) yields a power law for
the RTD,
\begin{eqnarray}
\psi_c(\tau)\propto \tau^{-\alpha/2}.
\end{eqnarray}
Furthermore, for intermediate times $ \tau_0 \ll \tau
\ll\tau_D$, Eq. (\ref{asym1direct}) yields another power law,
reading
\begin{eqnarray}\label{asym3}
\psi_c(\tau)\propto
\tau^{-\beta}
\end{eqnarray}
with $\beta=2-\alpha/2$. For $\alpha=1$ such an
intermediate power law with the
slope $\beta=3/2$ has been measured for a potassium ion channel in
\cite{Sansom}. Note that this particular power law exponent reflects
normal diffusion. Consequently, the origin of the intermediate power
law is principally not due to the anomalous diffusion behavior. Our
theory smoothly reproduces the intermediate power law associated
with normal diffusion in the limit $\alpha\to 1$.

Other  power law
features were also measured in experiments, for example,
for a gramicidin channel with the slope
$\beta\approx 1.7$ \cite{gramicidin}. This corresponds
to an intermediate power law in Eq. (\ref{asym3}) with $\alpha\approx 0.6$.
Moreover,  power law exponents with $\beta>2$ are also measured in experiments
\cite{Gorz} analyzed in Ref. \cite{Mercik} from a pure
phenomenological perspective without clarifying a tentative
mechanism. Our model  can as well capture  such anomalous power
laws which cannot be explained within the intermediate power law
asymptotics (\ref{asym3}).

Indeed, let us assume that $\tau_D$ is sufficiently small and to
consider the asymptotic behavior $\tau\to \infty$. The corresponding
asymptotics can be deduced from the behavior of $\tilde \psi_c(s)$
at small $s$. For $s\to 0$, Eq. (\ref{mainres}) yields,
\begin{eqnarray}
\tilde \psi_c(s) \approx 1-s
\langle \tau_c\rangle [1-(s\tau_D)^{\alpha}/3].
\end{eqnarray}
From this, by way of $\tilde \Phi_c(s)=(1-\tilde \psi_c(s))/s\approx
\langle \tau_c\rangle (1-(s\tau_D)^\alpha/3)$ for $s\to 0$, it follows
\cite{LS} that
$\Phi_c(\tau)\propto 1/\tau^{1+\alpha}$ for $\tau\to\infty$. This
renders then a power law (\ref{asym3}) with
$\beta=2+\alpha$ for large $\tau$. This asymptotic power law with $\beta>2$ is a
manifestation of the anomalously slow conformational diffusion in
a space domain of {\it finite} size. It
replaces an exponential asymptotic behavior of $\psi_c(\tau)$ for
$\tau>\tau_D$ in the case of normal diffusion
\cite{Millhauser,PNAS02,Physica03}.
\section{Application to gating dynamics of a locust potassium ion channel}

Our fractional diffusion model can be used to describe  the
rather complex gating behavior observed for a locust potassium ion
channel \cite{Mercik}. This ion channel exhibits experimentally a
Pareto law in its gating kinetics, $\psi_c(\tau)=a/(b+\tau)^\beta$
with $\beta\approx 2.24\pm 0.06$. The corresponding autocorrelation
function, however, seems to exhibit three different interchanging
power laws \cite{Mercik}. These features are compatible with our
model. Within a two-state reduction we are dealing with an
alternating renewal process \cite{Cox}. Its (Laplace-transformed)
normalized autocorrelation function reads \cite{Stratonovich,GH03}
\begin{eqnarray}\label{laplace-corr}
\tilde k(s)=\frac{1}{s}-\left(\frac{1}{\langle \tau_o\rangle} +
\frac{1}{\langle \tau_c\rangle}\right)\frac{1}{s^2}
\frac{\left(1-\tilde \psi_o(s)\right)
\left(1-\tilde \psi_c(s)\right)}{\left(1-\tilde \psi_o(s)\tilde
\psi_c(s)\right)}.
\end{eqnarray}
For our case under study this yields
\begin{eqnarray}\label{thismodel}
\tilde k(s)=\frac{1}{s}\frac{f_{\alpha}(s\tau_D)+s\langle \tau_c\rangle}
{1+\frac{\langle \tau_c\rangle}{\langle \tau_o\rangle}+f_{\alpha}(s\tau_D)+
s\langle \tau_c\rangle },
\end{eqnarray}
where $f_{\alpha}(z):=1/g_{\alpha}(z)-1$. Note that $f_{\alpha}(0)=0$
and for $\tau_D=0$ the inversion of (\ref{thismodel}) yields the
Markovian result $k(t)=\exp[-(k_0+k_c)t]$. Moreover, the analytical
expression (\ref{thismodel})
allows one to study the asymptotics of $k(t)$ at $t\to \infty$.
 Namely, from $f_{\alpha}(s\tau_D)
\approx (s\tau_D)^\alpha/3$ it follows that
$\tilde k(s)\propto s^{\alpha-1}$ at $s\to 0$. By virtue of a
Tauberian theorem \cite{Doetsch}, this latter
result readily yields,
\begin{eqnarray}\label{new4}
k(t)\propto t^{-\alpha}\;.
\end{eqnarray}
This power law feature agrees
well with the experiment which shows asymptotically (\ref{new4})
with $\alpha=0.28\pm 0.1$.
Furthermore, an intermediate asymptotics of
$k(\tau)$ can be obtained by studying the limit of very large $\tau_D$.
Using the scaling $\tilde s:=s\langle \tau_c\rangle$
and the limit of very large values
$y:=\tau_D/\langle \tau_c\rangle\gg 1$
such that $\tilde s \ll 1$ is allowed for, whereas still
$\tilde s y\gg 1$, Eq. (\ref{thismodel}) can formally
be approximated by
\begin{eqnarray}\label{largetauD}
\tilde k(s)= \frac{s^{\alpha/2-1}}{s^{\alpha/2}+r^{\alpha/2}}\;,
\end{eqnarray}
with $r=\tau_D^{-1}
(1+\langle \tau_c\rangle/\langle \tau_o\rangle)^{2/\alpha}$.
The formal inversion of
Eq. (\ref{largetauD}) yields $k(t)=E_{\alpha/2}[-(rt)^{\alpha/2}]$.
This in turn yields an intermediate asymptotics
$k(\tau)\propto \tau^{-\alpha/2}$ within
$\langle \tau_c\rangle \ll \tau\ll \tau_D$.
Indeed, the analysis  of experimental data in \cite{Mercik}
reveals such an
intermediate asymptotics $k(\tau)\propto \tau^{-0.14\pm 0.02}$.
\begin{figure}
\begin{center}
\includegraphics[width=.48\textwidth]{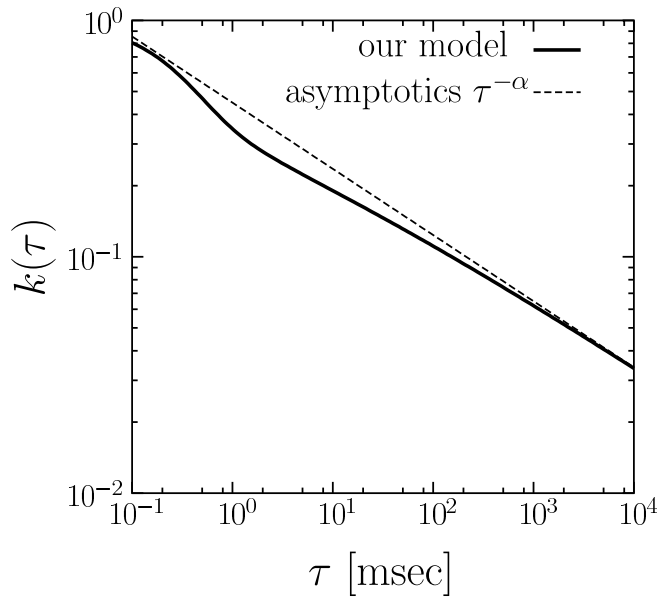}
\caption{Normalized autocorrelation function of conductance
fluctuations. Numerical inversion of Eq. (\ref{thismodel})
is done with the (improved) Stehfest algorithm \cite{Stehfest} for the
following set of parameters: $\langle \tau_c\rangle=0.84$ msec,
$\langle \tau_o\rangle=0.79$ msec \cite{Mercik} and assumed
$\tau_D=100$ msec and $\alpha=0.28$.}
\end{center}
\end{figure}
The numerical inversion of $\tilde k(s)$
in Fig. 2 displays  three different power law regimes
in qualitative agreement with the experimental data.
Only one of these
power laws
-- the long-time asymptotical one -- seems, however, to present
a true power law asymptotics.
The intermediate power law in Fig. 2
does not agree numerically with the experimental
one in Ref. \cite{Mercik}. Nevertheless, the experimental data agree --
surprisingly enough -- with the intermediate asymptotics
obtained above in the
limit $\tau_D\to \infty$.

Furthermore, the numerical inversion of $\tilde \psi_c(s)$
 in Fig. 3 can be fitted by Pareto law with $\beta\approx 2.24$.
 The discrepancy between $\beta-2\approx 0.24$ and $\alpha=0.28$
 is due to the experimental restrictions on the maximal time intervals
 measured. The actual power law asymptotics for $\tau\to \infty$
 in Fig. 3 is $\Phi_c(\tau)\propto \tau^{-1.28}$. This long time asymptotic
 regime is not yet attained in Fig. 3 which
 instead closely agrees with $\Phi_c(\tau)\propto \tau^{-1.24}$ , see in 
 Fig. 3, giving an apparent
 agreement with the experimental data.
\begin{figure}
\begin{center}
\includegraphics[width=.48\textwidth]{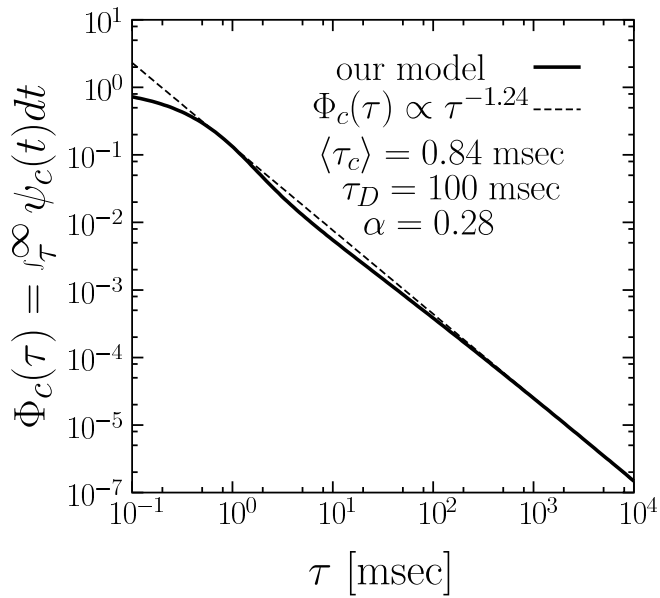}
\caption{Survival probability of the closed state for the studied
model. The set of parameters is the same as in Fig.2.}
\end{center}
\end{figure}
In view of our few elementary model assumptions, the agreement between
theory and the experimental data \cite{Gorz}
analyzed in Ref. \cite{Mercik} is striking indeed.

Our  fractional diffusion scheme is not expected
to describe the experimental facts quantitatively in all  details.
In particular, it predicts
that the low-frequency part of the spectral power $S(f)$ of
ion current fluctuations of the locust BK potassium ion channel
corresponds to $1/f^{\gamma}$ noise with $\gamma=1-\alpha\approx 0.72$
\cite{LowenTeich,GH03}. The experiment \cite{Siwy} indeed reveals
$1/f^{\gamma}$ noise with $\gamma$ close but to unity, $\gamma\approx 1$.
The reason for this discrepancy is not resolved. The asymptotic behavior of the
autocorrelation function in Ref. \cite{Mercik} and the behavior
of the low-frequency part of the spectrum in
Ref. \cite{Siwy} are certainly at odds.
A possible reason could be nonstationarity of the given 
current recordings. Nevertheless, the qualitative agreement, i.e.,
the principal theoretical prediction and the measurement of 
$1/f^{\gamma}$ noise,
is comforting.

Moreover, the durations of residence time intervals in  open
and closed states can be  correlated. Such correlations
can be induced by stochastic binding of calcium ions which regulate
the gating dynamics of large conductance potassium ion channels.
To account for such correlation effects, our model principally can
be generalized  in the same
spirit like the original diffusion model has been generalized
to include ligand binding effects  \cite{OswaldMillhauserCarter}.
This generalization is left but for a future study.

\section{Discussion and Conclusion}

The gating dynamics of  protein ion channels in biological membranes
is governed by a  conformational dynamics on a very complex energy
landscape with a huge number degrees of freedom. This
multidimensional energy landscape can possess deep energy wells (as
compared with the thermal energy $k_BT$) which are separated by
potential barriers. In addition, there can exist an underlying energy
valley network connecting these wells which results in an energy
quasi-degeneracy. The traditional discrete state approach to the
gating dynamics pursued by the community of molecular physiologists
presents an abstraction to this complexity: it has its focus on the
fact of deep potential wells being separated by high energy
barriers. The energy quasi-degeneracy of potential wells
  enters the theory as  an entropic contribution to
 the corresponding free energies after reduction of the multidimensional
reality to low-dimensional models (possessing  a few discrete states
only). This traditional approach has proven useful over the past
years and it serves as a serviceable working tool for
 the analysis of the experimental data. This approach is,
 however, not able to capture the physical origin of such
complexity features as the presence of power
law distributions of the residence time intervals, the slow decay of
the autocorrelations of fluctuations or the presence of
$1/f^{\gamma}$ noise feature in the power spectrum of fluctuations
of several ion channel types, to name but a few. Experiments have
demonstrated \cite{Bezrukov} that the $1/f^{\gamma}$ noise  is due
to the conformational transitions among different conductance
states. In particular,  the ion current is free of $1/f^{\gamma}$
noise in a frozen conductance substate of the ion channel.
Therefore, the $1/f^{\gamma}$ noise originates due to fluctuations
among experimentally distinguishable  substates \cite{Bezrukov}.
  These features reveal
unambiguously the non-Markovian character of the {\it observed}
``on-off'' ion current fluctuations \cite{Liebovitch,West,Fulinski}.
The diffusion models of ion channel gating
\cite{Millhauser,Lauger,Condat,Levitt,NadlerStein,Shirokov,PNAS02,
Physica03} present another, complementary abstraction of the actual
dynamics. These approaches attempt to capture the spatial structures
of the potential minima and the associated dynamics, and/or the
corrugated and hierarchical features of the real multidimensional
conformation landscape after performing the reduction
 to a reaction coordinate picture \cite{Frauenfelder}.
It is physically likely that the ion channel protein can become
temporarily trapped in some  domains of its intrinsic conformational
landscape  from which it can escape by activated jumps among those
states. Due to a complicated structure of such traps the
corresponding residence time distribution can possess a divergent,
-- or from a practical point of view --
 a very large  first moment. This in turn
gives rise to an anomalous conformation diffusion within the chosen
reduced reaction coordinate description. Our scheme in terms of a
fractional, continuous diffusion model, being complemented with
appropriate boundary conditions properly  accounts for such
complexity.

As demonstrated theoretically and exemplified with the gating of a
BK locust potassium ion channel our fractional diffusion theory
presents a powerful approach to describe these various observed
power law characteristics of the underlying gating dynamics.

\acknowledgments
This work has been supported by the Deutsche
Forschungsgemeinschaft via the collaborative research centre, {\em Manipulation of matter on the nanoscale},
 SFB-486, project A-10.

\appendix

\section{Solution of the boundary-value problem}

The Laplace-transformed probability $\tilde P(x,s):=\int_0^{\infty}e^{-st}
P(x,t)dt$, Eq. (\ref{fracdif}) reads:
\begin{eqnarray}\label{A1}
s\tilde P(x,s)-\delta(x-x_0)=K_{\alpha}s^{1-\alpha}
\frac{d^2 \tilde P(x,s)}{d x^2}
\end{eqnarray}
with $-L<x_0<0$. Note that the limit $x_0\to 0_{-}$ will be taken at
the very end of calculation. The corresponding Laplace-transformed boundary conditions
assume the form
\begin{eqnarray}
\frac{d \tilde P(x,s)}{dx}\Big|_{x=-L}& = &0, \label{A2}\\
s^{1-\alpha}\frac{d \tilde P(x,s)}{dx}\Big|_{x=0}& = &
-\frac{L}{K_{\alpha}\langle \tau_c\rangle}\tilde P(0,s)\;. \label{A3}
\end{eqnarray}
The challenge is thus the solution of  the boundary-value problem  (\ref{A1})-(\ref{A3}).
Towards this goal we consider separately the solution in the
domains, $-L<x<x_0$,
\begin{eqnarray}\label{A4}
\tilde P_1(x,s)=A_1 \exp\Big (\sqrt{\frac{s^\alpha}{K_{\alpha}}}x\Big)+
B_1 \exp\Big (-\sqrt{\frac{s^\alpha}{K_{\alpha}}}x \Big),
\end{eqnarray}
and  $x_0<x<0$,
\begin{eqnarray}\label{A5}
\tilde P_2(x,s)=A_2 \exp\Big (\sqrt{\frac{s^\alpha}{K_{\alpha}}}x\Big)+
B_2 \exp\Big (-\sqrt{\frac{s^\alpha}{K_{\alpha}}}x \Big).
\end{eqnarray}
At $x=x_0$, the solution is continuous,
\begin{eqnarray}\label{A6}
\tilde P_1(x_0,s)= \tilde P_2(x_0,s).
\end{eqnarray}
The first derivative $d \tilde P(x,s)/dx$, however,
experiences a jump.  This can  readily  be
seen upon integrating Eq. (\ref{A1}) in an infinitesimally small
neighborhood of $x=x_0$.
Thus,
\begin{eqnarray}\label{A7}
K_{\alpha} s^{1-\alpha}\Big [\frac{d\tilde P_2(x,s)}{dx}-
\frac{d\tilde P_1(x,s)}{dx}  \Big]_{x=x_0} =-1\;.
\end{eqnarray}
The coefficients $A_{1,2}$ and $B_{1,2}$ are determined by substitution
of (\ref{A4}) and (\ref{A5}) into Eqs. (\ref{A2}), (\ref{A3})
(\ref{A6}) and (\ref{A7}). Thereby, the proposed objective is exactly solved.
The integration of the solution (\ref{A4}) from $x=-L$ to $x=0$ ($x_0\to 0_{-}$)
yields the Laplace-transformed survival probability $\tilde \Phi_c(s)$;
the corresponding RTD (\ref{mainres})
follows as $\tilde \psi_c(s)=1-s\tilde \Phi_c(s)$.



\begin{thebibliography}{100}
\bibitem{remark}See also \cite{Goychuk04} where the GME of
Ref. \cite{Kenkre73} and its generalization by Burshtein, Zharikov and
Temkin  \cite{BurstZharTem} (used here) are reproduced from the most
general GME of CTRW with uncorrelated jumps, see Eq. (A8) in
\cite{Goychuk04}.
\end{thebibliography}
\end{document}